\documentclass{epl}

\title{Temporal Diffusion}

\author{Jean Pierre Boon \inst{1} \and Patrick Grosfils\inst{1} 
\and James F. Lutsko \inst{1}}
\institute{
  \inst{1} Center for Nonlinear Phenomena and Complex Systems\
       Universit\'e Libre de Bruxelles, 1050 - Bruxelles, Belgium
       (E-mail: {\tt jpboon@ulb.ac.be})
}

\begin{document}

\maketitle

\begin{abstract}
We consider the general problem of the first passage distribution 
of particles whose displacements are subject to time delays. 
We show that this problem gives rise to a 
\emph{propagation-dispersion equation} which is obtained as the 
continuous limit of the exact microscopic \emph{first visit equation}. 
The propagation-dispersion equation should be contrasted with the
advection-diffusion equation as the roles of space and time are 
reversed, hence the name \emph{temporal diffusion}, which is a 
generic behavior encountered in an important class of systems.
\end{abstract}

One of the fundamental physical paradigms, applicable to a wide variety of
physical processes, is that of spatial diffusion. The text-book example is
of a random walker on a one-dimensional lattice (see, e.g. \cite{feller}). 
At each tick of the clock, the walker takes a step either left or right, the
direction chosen randomly with equal probabilities, and one asks what is the
probability that the walker will be at a given position after a given time.
If the walker starts at a known point, the answer is a binomial distribution
which, in the continuum limit, becomes a Gaussian. The variance of the
Gaussian grows with time so that, the localization of the walker decreases
and we say that the walker disperses. If the probability for the walker to
step in one direction is greater than that for the opposite direction, then
the walker propagates in the direction of higher probability and will
eventually visit each site of the lattice in that direction. The typical 
diffusive behavior is then manifested in the continuum limit as a Gaussian 
about a most-likely position which moves at a constant velocity. However, 
there are a number of situations in which it is more natural to ask how 
long it will take to reach a given point - more precisely for a stochastic 
process, what is the distribution of times taken to reach that point. 
Everyday examples involve processes in which the goal is to arrive at a 
given point as quickly as possible:\ for example, a marathon (wherein we 
ask for the distribution of finishing times), certain financial instruments, 
such as stock options (wherein we ask for the distributions of times needed 
for an asset to reach a certain value), traffic-flow (wherein we ask for
the distribution of arrival times at destination), and packet transport over
the internet. A more technical example, which inspired the present work, is
the behavior of certain cellular automata which model the motion of a
particle on a substrate of scatterers (in 1 or 2 dimensions) where, for
certain types of scatterers, the particle ends up propagating along a
particular channel \cite{ant, cohen, grosfils}, and, again, the 
first-passage time is the physical quantity of interest. 
In this letter, we show that this general problem gives rise to a 
\emph{propagation-dispersion equation}, much like the biased random walker, 
with the important difference that the roles of space and time are
reversed: the distribution of first passage times is Gaussian in the time
variable with a variance that grows with increasing distance from the origin. 
In analogy with spatial diffusion that occurs in ordinary diffusive 
phenomena, we call this generic behavior \emph{temporal diffusion}.

Consider a walker on a one-dimensional lattice and let $\widehat{f}(t/\delta
t;r/\delta r)$ be the probability that it takes $t/\delta t$ time steps to
reach the lattice position $r/\delta r$, given that the walker is at the
origin at time $t=0$. Whatever the microscopic dynamics, we assume that we
are given, or can work out, the set of probabilities $\{p_{j}(r)\}_{j=1}^{%
\infty }$ that the time between the first visit of the lattice site $%
r/\delta r$ and the first visit of the next position, $r/\delta r+1$, is $%
j\delta t$. Conceptually, these represent the probabilities of various
waiting times from the first visit of lattice site $r/\delta r$ until the
first visit to $r/\delta r+1$, i.e. the distribution of single-step waiting
times. It is then clear that the probability that it takes the walker time 
$t$ to reach the lattice site $r+\delta r$ is equal to the probability that 
it takes time $t$ to reach lattice site $r$ and that the waiting time is zero,
plus the probability that it takes time $t-\delta t$ and that the waiting
time is $\delta t$, plus ...so that the master equation is \cite{boon}%
\begin{equation}
\widehat{f}(t/\delta t;r/\delta r+1)=\sum_{j=0}^{\infty }p_{j}(r)
\widehat{f}(t/\delta t-j;r/\delta r)\,.  
\label{1}
\end{equation}%
Rather than solve this equation directly, we appeal to a more intuitive 
picture. Let $\widehat{t}_{l}$ be the time that transpires from the first 
visit of lattice site $l$ to the first visit of lattice site $l+1$. 
The probability distribution for this discrete stochastic variable is 
precisely the set of probabilities $\{p_{j}(l\delta r)\}_{j=0}^{\infty }$ 
defined above. Next, define $\widehat{T}_{l+1}=\sum_{i=0}^{l}\widehat{t}_{i}$ 
which is the total time required to reach the lattice site $l+1$: 
the probability that $\widehat{T}_{l+1}$ takes on the value $t$ is precisely 
$f(t/\delta t;r/\delta r+1)$. Since each of the elementary stochastic 
processes $\widehat{t}_{i}$ is independent, the distribution of 
$\widehat{T}_{l+1}$ in the limit of large $l$ is immediately known, by 
application of the central limit theorem, to be%
\begin{equation}
f(t,r)=\lim_{r/\delta r>>1}\frac{1}{\delta t} 
\widehat{f}(t/\delta t;r/\delta r+1)=\sqrt{\frac{1}{2\pi 
\sigma^2 (r)%
}}\exp \left( -\frac{\left( t-\tau \left( r\right) \right) ^{2}}{%
2\sigma^2 (r)}\right)\,,   
\label{2}
\end{equation}%
where the most likely time is%
\begin{equation}
\tau \left( r\right) =\sum_{k=1}^{r/\delta r}\left\langle \widehat{t}%
_{k-1}\right\rangle =\delta t\sum_{k=1}^{r/\delta r}\sum_{j=0}^{\infty
}j p_{j}(\left( k-1\right) \delta r)\,,  
\label{3}
\end{equation}%
and the width of the distribution is%
\begin{eqnarray}
\sigma^2 (r)
&=&\sum_{k=1}^{r/\delta r}\left( \left\langle 
\widehat{t}_{k-1}^{2}\right\rangle -\left\langle \widehat{t}_{k-1}
\right\rangle^{2}\right) \nonumber \\
&=& \left( \delta t\right) ^{2}
\sum_{k=1}^{r/\delta r}\left[
\sum_{j=0}^{\infty }p_{j}(\left( k-1\right) \delta r)j^{2}-\left(
\sum_{j=0}^{\infty }p_{j}(\left( k-1\right) \delta r)j \right) ^{2}\right]\,. 
\label{4}
\end{eqnarray}%
(We note in passing that the exact solution to Eq.(\ref{1}) is a multinomial
distribution, as discussed in detail in \cite{archives}). If the
probability distribution of single-step waiting times is independent of
position, then $\tau (r)$ and $\sigma^2 (r)$ reduce respectively to 
\begin{eqnarray}
\tau \left( r\right)  &=&r\frac{\delta t}{\delta r}\sum_{j=0}^{\infty
}j p_{j} \,, 
\label{5a} \\
\sigma^2 (r) &=&r\frac{\left( \delta t\right) ^{2}}{\delta r}\left[
\sum_{j=0}^{\infty }j^{2}p_{j}-\left( \sum_{j=0}^{\infty }j p_{j}\right)^{2}%
\right] \,,  
\label{5b}
\end{eqnarray}%
which prompts us to define an inverse propagation speed as%
\begin{equation}
\frac{1}{c}= \frac{\tau\left( r\right)}{r} =  \sum_{j=0}^{\infty }j p_{j}
\frac{\delta t}{\delta r}\,,
\label{6}
\end{equation}%
and a temporal dispersion coefficient as%
\begin{equation}
\gamma = \frac{\sigma^2 (r)}{r} =  \left[ \sum_{j=0}^{\infty }j^{2}p_{j}-
\left( \sum_{j=0}^{\infty }j p_{j}\right)^{2} 
\right] \frac{\left( \delta t\right) ^{2}}{\delta r}\,. 
\label{7}
\end{equation}%
So the distribution of first-passage times is%
\begin{equation}
f(t,r)=\sqrt{\frac{1}{2\pi \gamma r}}%
\exp \left( -\frac{\left( t - r/c\right) ^{2}}{2\gamma r}\right) 
\label{8}
\end{equation}%
which is in precise analogy to the spatial distribution of a simple, biased
random walker: the most likely first-visit time grows linearly with
increasing distance and the width of the distribution grows as the
square-root of the distance.

The distributions given in Eqs.(\ref{2}) and (\ref{8}) are particular
solutions for the initial condition that the walker is localized at $r=0$
and $t=0$, or $f(t/\delta t;r/\delta r=0)=\delta (t)$. To complete our
description of the first-passage time problem, it is interesting to display
the continuum equivalent of the first passage equation, Eq.(\ref{1}), which
would govern the problem for all initial conditions. First, we notice that,
since the exact first-passage time equation is linear, the particular
solutions given above are the Green's functions for the general problem.
Explicitly, if $f(t/\delta t;r/\delta r=0)=f_{0}\left( t\right) $ then the
distribution for finite distances must be%
\begin{equation}
f(t,r)=\int_{-\infty }^{\infty }\sqrt{\frac{1}{2\pi \sigma^2 (r)}}%
\exp \left( -\frac{\left( t-t' -\tau\left( r\right) \right) ^{2}}{%
2\sigma^2 (r)}\right) f_{0}\left( t' \right) dt' \,,  
\label{9}
\end{equation}%
from which one finds the equation of motion, the
\emph{\ propagation-dispersion equation},%
\begin{equation}
\frac{\partial }{\partial r}f(t,r)+\frac{1}{c(r)}\frac{%
\partial }{\partial t}f(t,r)=\frac{1}{2}\gamma \left(
r\right) \frac{\partial ^{2}}{\partial t^{2}}f(t,r)
\label{10}
\end{equation}%
with 
\begin{eqnarray}
\frac{1}{c(r)} &=&\frac{\partial }{\partial r}\tau(r) \,, 
\label{11a} \\
\gamma \left( r\right)  &=&\frac{\partial }{\partial r}
\sigma^2 \left(r\right) \,.
\label{11b}  
\end{eqnarray}%
Equation (\ref{10}) can also be derived directly from Eq.(\ref{1}) 
by means of a multi-scale expansion (see \cite{archives}).

As a first example, consider a biased random-walker in one dimension.\ At
each tick of the clock, the walker moves to the right with probability $p$
and to the left with probability $q=1-p$. If the walker is at some
particular lattice site, say $l$, then the probability that it moves to $l+1$
with the next step is $p$, the probability that it takes three steps to move
to $l+1$ is $p(pq)$ since the probability of moving left, to $l-1$, and back
is $pq$ and in general, the probability that it takes $2m+1$ steps to reach 
$l+1$ is evidently of the form $p_{2m+1}=a_{m}\left( pq\right) ^{m}p$, where 
$a_{m}$ is a combinatorial factor independent of $p$, since a delay of $2m$
ticks, starting and ending at lattice site $l$, requires some combination of 
$m$ steps to the left and $m$ steps to the right. (Obviously, the probability 
to reach $l+1$ in an even number of steps is zero.) If $p>q$, we expect that 
the walker must eventually reach the next lattice site to the right so that 
$\sum_{m=0}^{\infty }a_{m}\left( pq\right) ^{m}p=1$. Introducing $y=p(1-p)$,
one has $p=\frac{1}{2}\left( 1\pm \sqrt{1-4y}\right) $ with the
positive sign appropriate for $p>1/2$ and the negative sign otherwise; 
the normalization condition can then be written as (for $p>1/2$)
\begin{equation}
\sum_{m=0}^{\infty }a_{m}y^{m+1}=\frac{1}{2}\left( 1-\sqrt{1-4y}\right) \,,  
\label{12a}
\end{equation}%
and expansion of the right hand side gives $a_{m}=\frac{\left( 2m\right) !}{%
(m+1)!m!}$. \ Then, having the single-step waiting-time probabilities, and
noting that
\begin{equation}
\sum_{m=0}^{\infty }\left(\begin{array}{c}2m+1\\m\end{array}\right)
p^{m+1}q^{m}  = (p-q)^{-1} \,,  
\label{12b}
\end{equation}%
the first and second moments of the waiting times can be evaluated to give 
\begin{eqnarray}
\frac{1}{c} &=&\frac{\delta t}{\delta r}\frac{1}{\left( p-q\right)}\,,
\label{13a} \\
\gamma  &=&\frac{(\delta t)^{2}}{\delta r}\frac{4pq}{\left( p-q\right)^{3}} \,,
\label{13b}
\end{eqnarray}%
and the distribution of first passage times is given by (\ref{8}). The
continuous limit, in which both $\delta r$ and $\delta t$ go to zero, 
gives finite results for both the propagation speed and the dispersion
coefficient only if we simultaneously require that $p-q$ go to zero (just as 
in the usual discussion of the spatial-diffusion of the biased random walker
\cite{feller}). Writing $\delta r\rightarrow \epsilon \delta r_{0}$, $\delta
t\rightarrow \epsilon ^{\alpha }\delta t_{0}$ and $p-q\rightarrow k\epsilon
^{\beta }$, we find that 
\begin{eqnarray}
\frac{1}{c} &\rightarrow&
\frac{\delta t_{0}}{k\delta r_{0}}\,\epsilon ^{\alpha -\beta -1} \,,
\label{14a} \\ 
\gamma  &\rightarrow&  
\frac{(1-k^2 \epsilon ^{2\beta})(\delta t_{0})^{2}}{k^{3}\delta r_{0}}\,
\epsilon ^{2\alpha-1-3\beta}\,,  
\label{14b}
\end{eqnarray}%
which are finite provided that $\alpha =2$ and $\beta =1$. This scaling is
identical to that used to obtain the spatially-diffusive limit of the biased
random walker: the propagation speeds are identical, but the dispersion
coefficients are quite different, because in spatial diffusion, the
diffusion coefficient is independent of the scaling of the probabilities 
($D \rightarrow  (\delta r_{0})^{2}/\delta t_{0}$). The temporal 
dispersion coefficient is $\gamma = D/c^{3}$, which is also what one obtains 
by performing a change of variables ($r \rightarrow ct, t \rightarrow r/c$)
in the advection-diffusion equation.

A more complex example is provided by a walker moving on a lattice with
scatterers inducing time delays at some (or all) lattice sites, a model 
which may serve as a paradigm for various processes such as signal 
propagation in computer networks \cite{internet}, traffic flows 
\cite{traffic}, and evolutionary dynamics \cite{ant}. Of particular 
interest are models in which the properties of the scatterers also change 
with each scattering process. As a simple example, consider a particle 
moving on a one-dimensional lattice which has scatterers at each
lattice site. The scatterers can be in one of two states characterized
by  their ``spin'' which can take on the values ''up'' and ``down''. 
When a particle moves to a lattice site with spin up, nothing happens to 
it whereas at a spin down site, its velocity is reversed. In both cases,
the spin of the lattice site is reversed. This model was solved in
\cite{grosfils} where it was shown that, for any initial distribution
of spins (including the random distribution), the particle always ends 
up propagating in one direction or the other at a constant (average) rate. 
More surprisingly, the same result is obtained on a 2-D triangular lattice 
with an analogous dynamics (spin up (down) rotates the velocity by 
$+(-) 2\pi /3$) \cite{grosfils} and on the square lattice (when all 
scatterers are initially in the same state and rotate the velocity 
of the particle by $\pm \pi /2$ depending on the state). The latter model 
\,is known in the literature as ``Langton's ant''\cite{ant}. As discussed in
\cite{boon}, the distribution of first passage times 
\emph{along the direction of propagation} in all of these examples can be 
cast in the general form 
\begin{equation}
\widehat{f}(t/\delta t;\left( r+\rho \right) /\delta r)=\sum_{j=0}^{\infty}
{\tilde p}_{j}\widehat{f}(\left( t-\tau _{j}\right) /\delta t;r/\delta r) \,, 
\label{15}
\end{equation}%
where $\rho $ represents the elementary space increment along the
propagation strip and $\tau _{j}=\left( 1+ a\,j\right) b\,\delta t$ 
($a$ and $b$ are lattice-dependent integer constants \cite{boon}). 
Equation (\ref{15}) 
can be mapped onto the form given in Eq.(\ref{1}) by setting 
$p_{\left( 1+ a\,j \right)\,b}={\tilde p}_{j}$ and all other $p_{j}$'s$\,=0$. 
All of these models will therefore, in the limit of large spatial 
separations, be described by the propagation-dispersion equation  
and the distribution of first passage times will be Gaussian. 

In Fig.1 we give an illustration of temporal diffusive behavior in an
inhomogeneous system where we compare the theoretical large distance 
solution with simulation data.
The time delay probability is taken to be ${\tilde p}_{j=k}=p(1-p)^{k-1}$
where $p$ is linearly space dependent: $p\equiv p(r)=p(0) + \chi r/\delta r$.
The dots are the simulation data obtained by solving numerically the 
microscopic equation (\ref{15}) and the solid line is the solution of the
propagation-dispersion equation (\ref{10}). Notice that for space
independent probabilities, the width of the Gaussian in (\ref{8}) increases
like $\sqrt r$, i.e. in the example given in Fig.1 the width at 
$8000\,\delta r$ should be twice that at $2000\,\delta r$ . 
Here we observe essentially no change in the respective widths, 
a consequence of the spatial dependence of the waiting time probabilities.

\begin{figure}
\begin{center}
\resizebox{10cm}{5cm}{\includegraphics{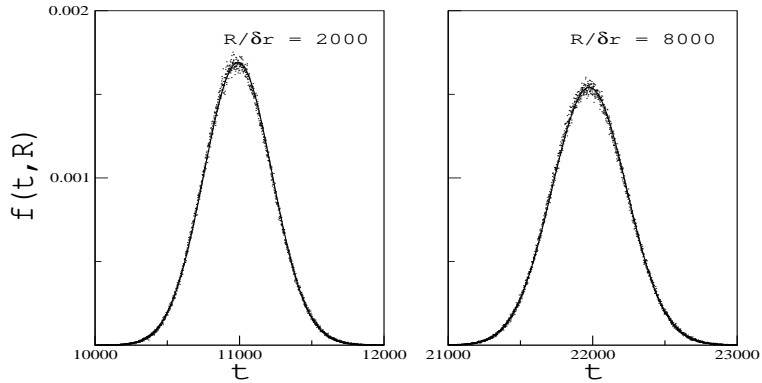}}
\end{center}
\caption{Temporal dispersion curve: $f(t, R)$ at 
$R/\delta r=2000$ and $8000$ (in lattice units). Simulation data (dots)
are obtained from the first visit equation (\ref{15}) with 
${\tilde p}_{j=k}=p(1-p)^{k-1}$ where 
$p\equiv p(r)=0.1 + 10^{-4} r/\delta r$. The theoretical curve (solid line)
is the Gaussian solution of the propagation-dispersion equation (\ref{10})
with $\tau(R)/\delta t=
R c^{-1}(R)(\delta t)^{-1}=\sum_{r=1}^{r=R}p^{-1}(r) = 10989$ 
and $21977$, and half-width $[2R \gamma(R)]^{1/2}(\delta t)^{-1}
=[\sum_{r=1}^{r=R}(1-p(r))p^{-2}(r)]^{1/2}= 334$ and $366$,
at $R/\delta r=2000$ and $8000$ respectively.}
\label{f.1}
\end{figure}

Recently, Buminovich and Khlabystova \cite{buni-khlabys} have studied 
models in which the scatterers only change state after multiple 
scattering events. In this case, the distribution of elementary waiting 
times becomes dependent on the lattice position, and the propagation speed 
and dispersion coefficient acquire a spatial dependence. Thus, while the 
distributions of first passage times are still Gaussian, they are not 
``diffusive'' in the usual sense since the inverse propagation speed and 
dispersion coefficient are not constants (Eqs.(\ref{10}-\ref{11b}) 
and reference\cite{archives}).

A more practical example is found in a recent report \cite{hulin} which
describes an experiment where small beads are dropped into a container 
filled with larger beads. The small beads, driven by gravity, percolate
through the array of larger beads, and their propagation is carefully measured.
Intuitively, one would imagine that the various deflections of the small bead 
as it collides with the larger ones would induce time delays in its downward 
propagation, in which case the distribution of arrival times would be given by 
Eq.(\ref{8}). This is indeed confirmed by the experimental results which 
exhibit the signature of a temporal-dispersion process (see Fig.9 in 
reference \cite{hulin}).

In summary, we have demonstrated that the distribution of first-visit times
of a particle propagating with stochastic time delays on a one dimensional
lattice, or the one-dimensional version of a multidimensional process, 
satisfies a temporal propagation-dispersion equation in the limit of large
separations. This generic behavior is analogous to the generic diffusive
behavior which describes the spatial distribution of the same process as a
function of time. For a simple biased random walker, the propagation speed
is the same in the temporal and spatial descriptions, while the temporal-
and spatial-dispersion coefficients differ.  The temporal-dispersion
description is relevant to a class of one- and two-dimensional cellular
automata in which mobile particles and fixed scatterers interact with one
another. Experiments in granular media also show behavior described by
temporal dispersivity. We have restricted our discussion to the case where
the variance of the elementary time-delay processes exist. There is an
interesting class of similar processes which are described by power-law 
distributions, e.g. $p_{j} \sim j^{-(1+\nu)}$, in which case, for
$0< \nu \leq 1$, the distribution appearing in the central limit 
theorem is no longer Gaussian (see, e.g., \cite{gnedenko}). 
This problem will be treated in a forthcoming publication.

\end{document}